\documentclass[%
 aip,
 amsmath,amssymb,
 reprint,%
]{revtex4-1}

\usepackage{graphicx}
\usepackage{dcolumn}
\usepackage{bm}
\usepackage{wasysym}
\usepackage{textcomp}

\usepackage[utf8]{inputenc}
\usepackage[T1]{fontenc}
\usepackage{mathptmx}
\usepackage{etoolbox}
\usepackage{xcolor}
\usepackage{soul}
\usepackage{mathtools}
\usepackage{subcaption}
\usepackage{svg}

\makeatletter
\def\@email#1#2{%
 \endgroup
 \patchcmd{\titleblock@produce}
  {\frontmatter@RRAPformat}
  {\frontmatter@RRAPformat{\produce@RRAP{*#1\href{mailto:#2}{#2}}}\frontmatter@RRAPformat}
  {}{}
}%
\makeatother
\begin{document}

\preprint{AIP/123-QED}

\title[Cryogenic Lagrangian Exploration Module]{The Cryogenic Lagrangian Exploration Module: a rotating cryostat for the study of quantum vortices in Helium II via particle seeding}

 \author{J. Vessaire}
\author{C. Peretti}%
 \altaffiliation[Now at: ]{National High Magnetic Field Laboratory, 1800 East Paul Dirac Drive, Tallahassee, Florida 32310, USA}
\author{F. Lorin}
 \author{E. Durozoy}%
 \author{G. Garde}
\author{P. Spathis}
\author{B. Chabaud}
\author{M. Gibert}
 \email{mathieu.gibert@neel.cnrs.fr}
 \affiliation{Univ. Grenoble Alpes, CNRS, Institut N\'eel, 38000 Grenoble, France
}%

\date{\today}

\begin{abstract}
The study of quantum vortex dynamics in He$_\mathrm{II}$ offers great potential for advancing quantum-fluid models. Bose-Einstein condensates, neutron stars, and even superconductors exhibit quantum vortices, whose interactions are crucial for dissipation in these systems. These vortices have quantized velocity circulation around their cores, which, in He$_\mathrm{II}$, are of atomic size. They have been observed indirectly, through methods such as second sound attenuation or electron bubble imprints on photosensitive materials. Over the past twenty years, decorating cryogenic flows with particles has become a powerful approach to studying these vortices. However, recent particle visualization experiments often face challenges with stability, initial conditions, stationarity, and reproducibility. Moreover, most dynamical analyses are performed in 2D, even though many flows are inherently 3D. We constructed a rotating cryostat with optical ports on an elongated square cupola to enable 2D2C, 2D3C, and 3D3C Lagrangian and Eulerian studies of rotating He$_\mathrm{II}$ flow. Using this setup, individual quantum vortices have been tracked with micron-sized particles, as demonstrated by Peretti \textit{et al.}, \textit{Sci. Adv.} {\bf{9}}, eadh2899 (2023). The cryostat and associated equipment—laser, cameras, sensors, and electronics—float on a 50~$\mu$m air cushion, allowing for precise control of the experiment’s physical parameters. The performance during rotation is discussed, along with details on particle injection.
\end{abstract}

\maketitle
\tableofcontents

\section{Introduction}

Over the past 117 years, since the first man-made droplet of helium \cite{Onnes}, many efforts have been made to explore its unique properties and develop models for liquid helium. These modeling attempts demonstrate that a single model encompassing the entire range of scales and thermodynamic properties (temperature and pressure) remains challenging to achieve. The study of the liquid Helium~II phase is fascinating because it exhibits non-classical properties, such as superfluidity  \cite{LandauV5,LandauV9}. Historically, Landau and Tisza proposed the two-fluid model. Still, it was limited by an unclear understanding of the two fluids and the absence of a crucial element: the quantum vortex \cite{Feynman}. Quantum vortices exhibit quantized velocity circulation around their cores, which are approximately atomic in size in He$_\mathrm{II}$. A model by Hall, Vinen, Bekarevich, and Khalatnikov \cite{Donnelly} provides a simplified explanation of the interactions among vortices, the superfluid, and the normal fluid. However, it does not capture the complex flow details relevant to quantum turbulence \cite{DonnellyVinen}. Since the early 2000s, Lagrangian and Eulerian approaches to fluid flow have gained prominence, driven by the development of high-speed cameras and increased data storage capacity \cite{Mordant,Ouellette,DonnellyPIV}. This progress led to G. Bewley's pioneering work, which visualized the first vortex lattice decorated with solid particles \cite{bewley}. Since then, various tracer particles have been used\cite{guo2014visualization}, including dihydrogen \cite{bewley2008particles}, dideuterium \cite{duda2017streaming,tang2023imaging}, hydrogen deuteride \cite{mantia_hd_particles} and air flakes \cite{fonda2016sub}, as well as nanometric tracers such as ion bubbles \cite{golov_electron} and $He_2^*$ excimers \cite{guo2010visualization}. Controlled production of dideuterium droplets has also been explored \cite{bret2023controlled}. Studies on particle-vortex interactions have been carried out both numerically \cite{giuriato2020trapped} and experimentally \cite{Prague}.  The investigation of rotating superfluid flows has been performed in $^3$He using demagnetization refrigerators \cite{Makinen}, and in He$_\mathrm{II}$ with helium bath cryostats \cite{Varga} and dilution refrigerators \cite{Golov_RSI}. Additionally, flow visualization systems have been developed \cite{Guo_RSI,goodwin_phd,Mantia_RSI}.

Hence, we propose an experiment specifically designed to feed flow-tracking data into the He$_\mathrm{II}$ models using the solid particle tracer technique. This approach aims to refine, exclude, or confirm the models. The Cryogenic Lagrangian Exploration Module (CryoLEM, named after the apparatus presented in Zimmermann et al. \cite{zimmermannLEM}) is a pressure/temperature-controlled rotating optical cryostat operating at saturated vapor pressure. This creates the canonical flow, which is the quantum vortex lattice formed by steady rotation \cite{Feynman}. The system is controlled and reproducible, providing a stable initial condition to explore the parameter space by varying temperature, rotational velocity, driving mechanism, or perturbation. Optical access enables particle seeding with solid dihydrogen or dideuterium, facilitating direct visualization data (see \ref{Optical diagnostics - Visualisation}), which is analyzed using Particle Tracking Velocimetry (PTV) and Particle Image Velocimetry (PIV). Additionally, a heating apparatus and channel allow He$_\mathrm{II}$ to be precisely driven by a canonical flow known as a counterflow, which is unique to this phase of matter. 

As described by Barenghi et al. \cite{Barenghi_review}: "Unfortunately, the only experimental information about rotating quantum turbulence is from a few studies of rotating thermal counterflow \cite{Peretti,Swanson} in helium, and thermal counterflow turbulence is an old problem still under investigation" (see Mastracci \textit{et al.}  \cite{mastracci2019characterizing} and \v{S}van\v{c}ara \textit{et al.} \cite{Prague}). The topic is gaining renewed interest as recent experiments, such as those by Varga \textit{et al.} \cite{Varga}, also provide new data. The apparatus presented in this paper has enabled tracking of individual quantum vortices forming a vortex lattice generated by rotating He$_\mathrm{II}$, as well as observing lattice destabilization and destruction when counterflow is applied \cite{Peretti}.

In subsection \ref{CryoLEM}, we introduce the CryoLEM by first describing its design and cryogenic performance. In subsection \ref{Rotation}, we discuss the rotating table and its alignment, followed by the configuration of cameras and optical ports used to visualize the dihydrogen or dideuterium particles, with the seeding process detailed in \ref{Optical diagnostics - Visualisation}. In \ref{Protocol}, we describe the typical experimental protocol. Finally, we cover the control and real-time data monitoring system in section \ref{Tools}, along with additional apparatus for various flow experiments.

\section{Experimental description}
In the following subsections, we detail the steps and methods needed to achieve our goal of studying low-temperature fluids through optical diagnostics.

\subsection{From Orange cryostat to CryoLEM : A cryostat for the visualization of rotating liquid helium flows} \label{CryoLEM}
\begin{figure*}[!htp]
    \centering
    \includegraphics[width=0.8\textwidth]{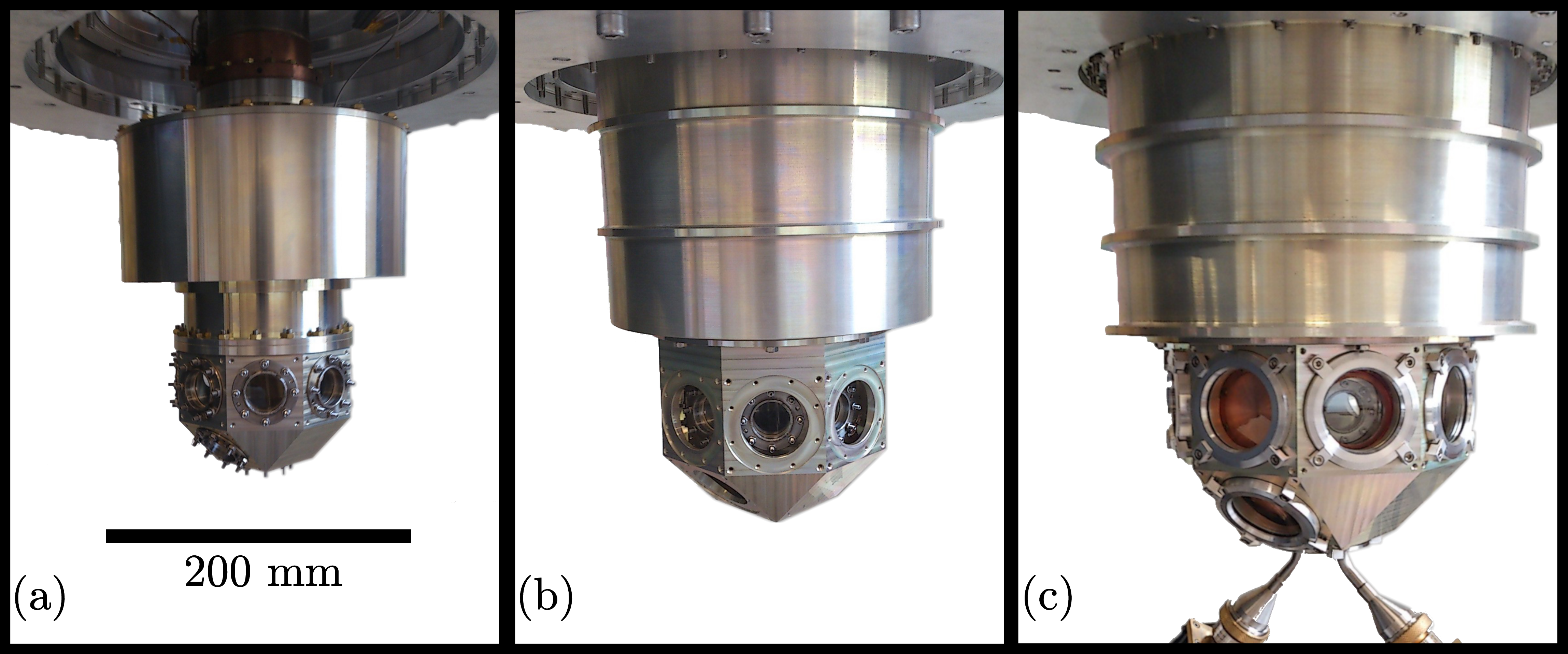}
    \caption{Experimental cell with optical ports. The helium reserve is located above the optical access ports, and thermometer data can be transmitted through a reconfigured optical port (here, the bottom port). (a) Inner cell, which contains the liquid helium. (b) $77$~K intermediate jacket in thermal contact with the nitrogen reservoirs. (c) $300$~K layer.}
    \label{expe_cell}
\end{figure*}
\begin{figure*}[!htbp]
    \centering
    \includegraphics[width=\textwidth]{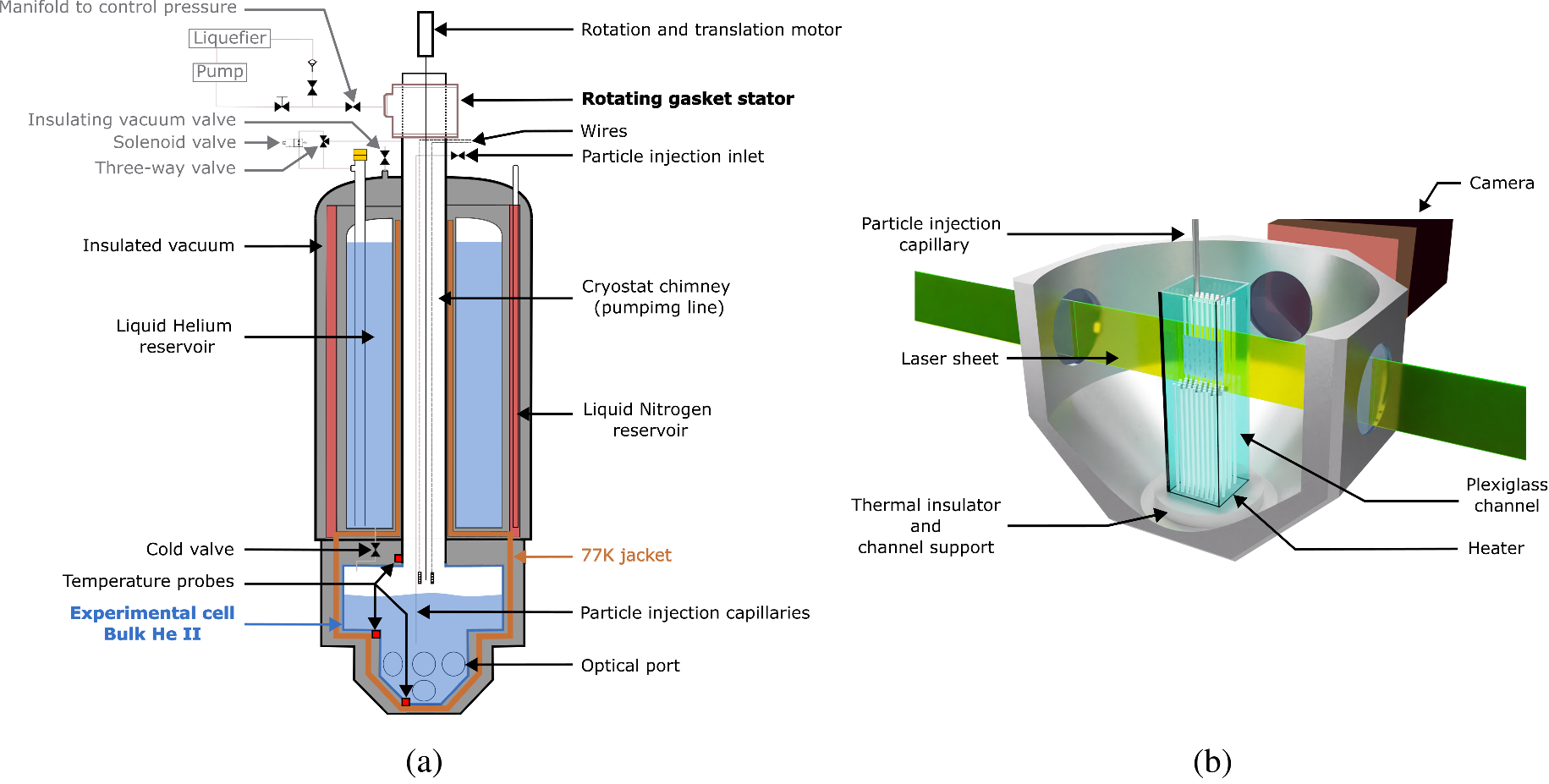}
    \caption{(a) Schematic view of the modified orange cryostat with the experimental cell. It has been modified by adding a cold valve, which allows filling the core of the cryostat with liquid helium from the reservoir, and a solenoid valve that permits releasing the evaporated helium gas from the reservoir to the top of the pumping line. The main modifications include the presence of the experimental cell with eight optical accesses at the bottom of the chimney, and the addition of particle injection capillaries that run from the top of the chimney to the middle of the experimental cell. In gray: pressure control apparatus.
    (b) Experimental setup zoomed in on the experimental cell. White rods schematically represent the vortex lattice present in rotating superfluid helium.}
    \label{SchemaCryostat}
\end{figure*}

Based on an Orange cryostat developed by the Institut Laue-Langevin (Grenoble) \cite{OrangeILL} and built by AS Scientific Products, the Cryogenic Lagrangian Exploration Module (CryoLEM) is an experimental device designed to address cryogenic challenges when generating model He$_\mathrm{I}$ and He$_\mathrm{II}$ rotating flows seeded with macroscopic particles (similar to classical hydrodynamics). The main modifications implemented are as follows: 
\begin{enumerate}
    \item Instead of placing a study sample at the bottom of the pumping chimney, an additional modular experimental cell with 8 optical accesses is added (see fig.~\ref{expe_cell}).
    \item The central chimney connects to the pumping line through an in-house-designed rotating gasket (see fig.~\ref{SchemaCryostat}a).
    \item The experimental setup and instruments are mounted on a rotating table.
\end{enumerate}

%

Similar to the working principle of Dewar flasks, the experimental optical cell is enclosed in two larger shields (see fig.~\ref{expe_cell} and fig.~\ref{SchemaCryostat}a), all separated by an insulating vacuum. The intermediate shield is in direct contact with the original nitrogen jacket of the Orange cryostat, thus thermally connected to $T \gtrsim 77$~K, therefore reducing radiative transfers. The geometry of the experimental cell can be divided into two parts: an upper cylinder (inner diameter \diameter 198~mm, height 92~mm) and a lower elongated square cupola (see fig.~\ref{expe_cell}), allowing flat optical access. The inner side of the lower part is axisymmetric (fig.~\ref{expe_cell}, \diameter 108mm) to prevent stirring caused by a polygonal shape \cite{Bach}. However, small recesses on the sides of the optical access are unavoidable due to their flat shape. The inner volume of the experimental cell is $3.9$~L.
The same eight optical accesses are placed on both the intermediate and external shields. The window configuration allows studying the fluid from different angles in 2D2C, 2D3C, and 3D3C\footnote{$x$D$y$C indicates $x$ observed dimensions and $y$ measured coordinates.}. For example, in 2D, the particles in the fluid can be illuminated by a laser sheet orthogonal to the rotational axis (horizontal) or containing it (vertical) — see fig.~\ref{SchemaCryostat}b for an experimental setup. In these cases, the camera optical axis should be orthogonal to the laser sheet, facing either the bottom window of the cell or a side window. In the configuration shown in fig.~\ref{SchemaCryostat}a, the bottom window is replaced with an access for sensor wires connecting to sensors on the outside of the experimental cell (like the temperature probes in fig.~\ref{SchemaCryostat}a). This access can be swapped with any other window if the bottom one needs to be used. Additionally, this window setup ensures there is always an exit window for the laser beam, which is essential to prevent heating the cryogenic fluid inside the cell with the laser.

The inner windows have a diameter of $25$~mm and are made of sapphire, which has good thermal conductivity even at low temperatures. This property ensures a uniform temperature across the window. Combined with its high transparency in the visible range, sapphire windows prevent the formation of counterflow sources that could be caused by absorption of laser light \cite{bib:Rousset2009_Cryogenics}.
The windows of the intermediate shield are made of KG3 — a technical glass transparent in the visible spectrum but with less than $1\%$ in the infrared domain \cite{bib:Melich2011_JLTP}, designed to block infrared radiative transfer from the 300~K exterior. Lastly, the outer windows are made of borosilicate glass, a standard material in vacuum applications. These are coated with an anti-reflective coating to ensure optimal optical performance. On the opposite end of the setup, two valves are installed at the top of the apparatus: one for manually filling the cell with liquid helium from the reservoir, and the other is a solenoid valve to release the evaporated helium from the reservoir into the top of the pumping line. Since the original $9$~L helium reservoir remains intact, its bottom is connected to the experimental cell via a capillary equipped with a "cold valve" (see fig.~\ref{SchemaCryostat}), which can be operated manually from the top of the cryostat at $300$~K. Having two separate helium reservoirs is practical because it reduces the volume of the experimental cell, allowing helium and particles to evaporate overnight and enabling experiments to start with a clean cryostat the next day. Moreover, during lengthy experiments, a significant amount of helium may evaporate. If needed, it is convenient to quickly refill the cell using a single operator by opening the cold valve.

\subsubsection{Temperature control and daily cooling down}
\begin{figure*}[ht!]
    \centering
    \includegraphics[width=0.9\textwidth]{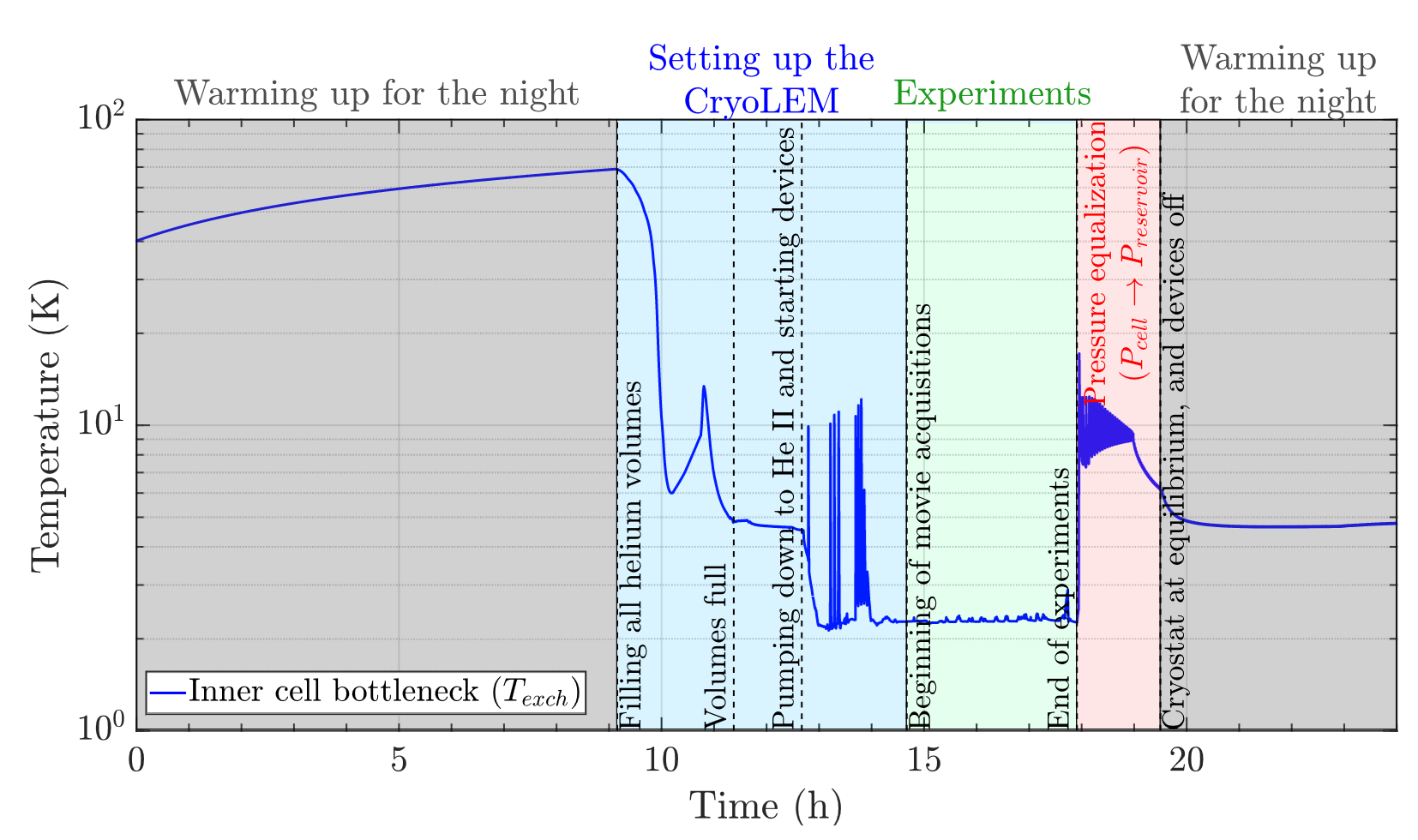}
    \caption{Typical temperature profile during experiments. In blue : temperature of the bottleneck between the cryostat chimney and the cell ($T_{exch}$). The fluctuations observed in the green "Experiments" region are due to particle injections ($300$~K gas) between each experimental run acquisition. Stable temperature is obtained thanks to pressure regulation.}
    \label{Tvst}
\end{figure*}

The helium bath temperature is controlled by adjusting the saturated vapor pressure (SVP) above the bath through regulated pumping. The pump used is a SOGEVAC SV300 B rotary vane oil pump. This controlled pumping is managed with a Leybold MR 50 Diaphragm Pressure Regulator integrated into the manifold (see fig.~\ref{SchemaCryostat}), which includes a control volume that sets the target pressure to pass through the connected pumping line. The control volume is connected to the pumping line via a Leybold DN KF16 ISO-KF variable leak valve, enabling precise pressure regulation (up to $10^{-4}$~bar) with pressure fluctuations of about $10^{-5}$~bar. Cernox\textregistered{} sensors mounted outside the experimental cell provide calibrated temperature readings, ensuring accurate monitoring of the helium bath. Pressure is measured using Keller PAA-41X pressure gauges.

The initial cooldown of the CryoLEM proceeds as follows: using PT100 thermometer placed on the external surface of the intermediate shield ($T_{N_2}$), we monitor the cooling process by conduction and radiation until the nitrogen jacket reaches $\sim 100$~K. The reservoir is then filled with liquid helium, a process taking about 30 minutes. Once this stage is achieved, the helium reservoir is cooled, while the experimental cell remains at a higher temperature. To cool it further, the cold valve is opened, allowing helium to flow from the reservoir. Since there are more metal parts to cool than in the original orange cryostat build, it takes roughly 5 to 6 hours to cool the experimental cell down to $4.2$~K. To speed up this process, controlled pumping is performed to reduce the pressure inside the experimental cell to $100$~mbar, increasing helium flow. This pressure control reduces the cooldown time to approximately 3 to 4 hours. Once the cell reaches $4.2$~K (T$_{He_{cell}} = 4.2$~K) and is filled with helium, the experimental runs can commence, as detailed in \ref{Protocol}. In the following days, the cooling time is greatly reduced to about 20 minutes because the cell temperature only rises to roughly $50$~K overnight, a temperature below which the thermal capacity of the cell components has already decreased significantly, and the helium reservoir still contains ready-to-use liquid. 

\subsubsection{Cryogenic performances}
Even if the cryostat is well insulated, some heat flux from the outside persists, gradually evaporating the fluid. We refill the nitrogen reservoir every morning and evening during experimental sessions, which is sufficient to keep its temperature consistently below $100$~K if the cell remains cold, as shown in fig.~\ref{Tvst}. Experiments demonstrate that the helium bath can be used for an entire afternoon without completely evaporating the liquid helium. To measure the evaporation rate, the surface of the liquid helium is observed for several minutes to accurately determine the lifespan of a superfluid helium bath in the CryoLEM. 

\begin{figure}[!htbp]
    \centering
    \includegraphics[width=1\linewidth]{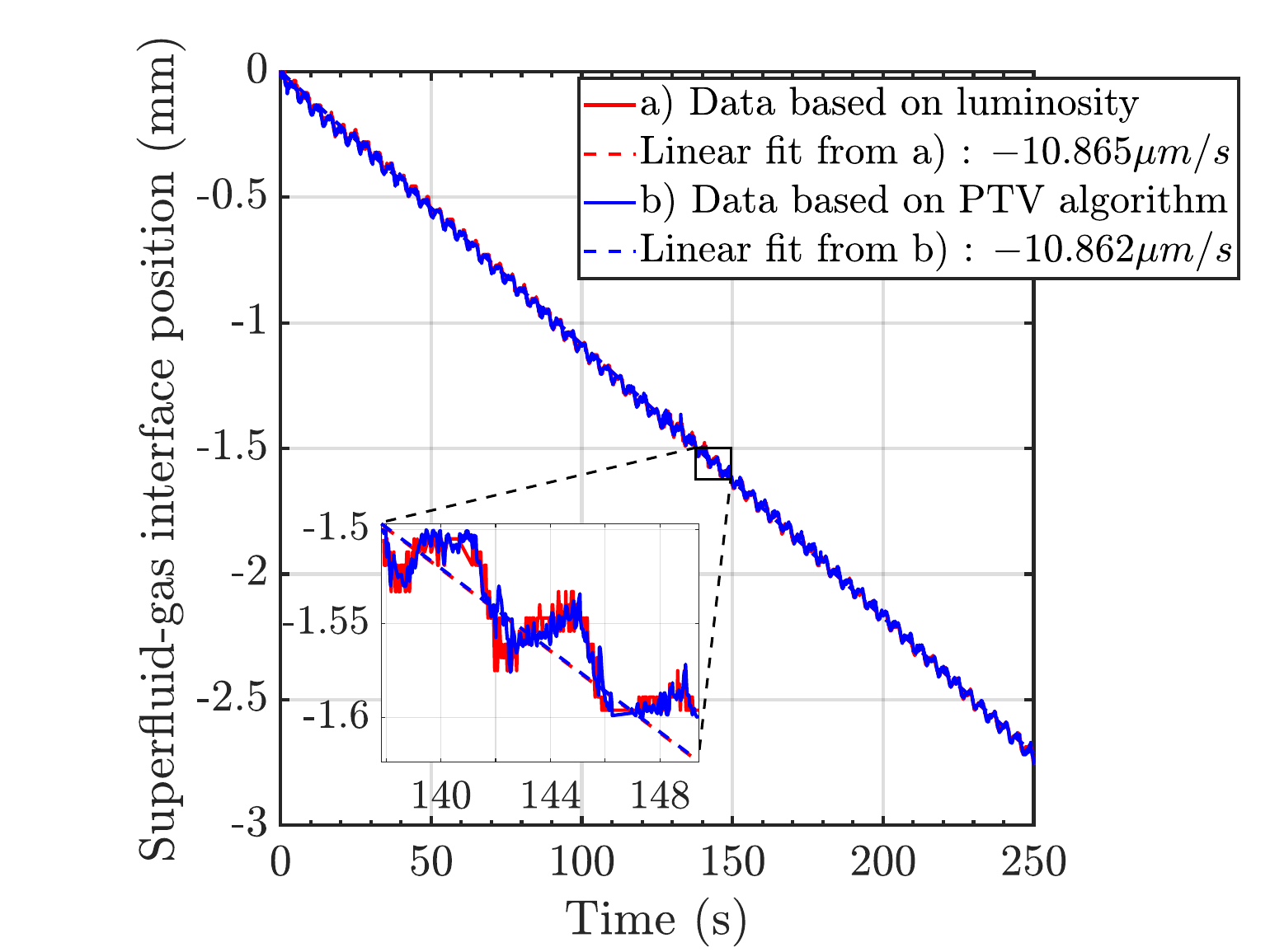}
    \caption{Superfluid-Gas interface position as a function of time is shown. A double plot compares the luminosity-based data and the PTV-based data, which overlap quite significantly as shown in the insert. Consequently, both methods yield a similar surface velocity of $v_{surf}\simeq -11$~$\mu$m.s$^{-1}$. A film is recorded during the experiment with a laser sheet orthogonal to the camera, capturing particles floating in the superfluid. The luminosity-based data is obtained by summing the pixel intensities horizontally and identifying the peak intensity. The PTV-based data is directly derived from the PTV tracks with the highest altitudes.} 
    \label{surfacebain}
\end{figure}

During this experiment, the superfluid is at 2.088~K (40~mbar), the cryostat is spinning at $15$~RPM, the laser power is set to $0.5$~W to visualize the deuterium particles, and the heater is set to $Q_{cell} = 0.2$~W to regulate the pressure with PID-controlled butterfly valves. This additional heat source is necessary because the evaporation caused by heat losses alone is too weak to be regulated. The fluid surface is already near the center of the image recorded by the camera when the measurement starts. After less than 7 minutes of recording, its surface is already outside the field of view. To track the fluid surface numerically, two approaches are tested:

\begin{itemize}
    \item Since the laser sheet appears brighter at the gas-superfluid interface, the intensity of each pixel row is summed, and the position of the maximum is used to identify the surface. This high luminosity may result from $D_2$ particles trapped by surface tension or from the low incident angle of the laser sheet, which causes reflections.
    \item A particle tracking velocimetry algorithm is applied to this video. For each frame, the highest vertical point of the detected particles is used to determine the surface position.
\end{itemize}

The results of both methods are shown in fig.~\ref{surfacebain}, and they are identical. We measure a constant surface velocity of $-10.9~\mu$m.s$^{-1}$, or $-39.1$~mm.h$^{-1}$. Let $\mathcal{V}_{He_\mathrm{II}}$ be the volume of the He$_\mathrm{II}$ bulk. Considering the diameter of 108~mm at the bottom of the cell, this corresponds to an evaporating flow rate of $\frac{d\mathcal{V}_{He_\mathrm{II}}}{dt} = 0.36$~L.h$^{-1} = 1.0 \cdot 10^{-7}$~m$^3$.s$^{-1}$. Using the latent heat of liquid helium ($\delta H^{l \rightarrow g}_{He} = 23.1$~kJ.kg$^{-1}$) and its density $(\rho_{He}(2.088 \text{~K}) = 145.8 \text{~kg.m}^{-3})$, the heat loss of the cryostat can be estimated:
\begin{equation*}
  \mathcal{Q}_{loss}=\Delta H_{He}^{l \rightarrow g} \rho_{He}  \dfrac{d\mathcal{V}_{He_\mathrm{II}}}{dt} - Q_{cell} = 0.14 \text{~W}
\end{equation*}
This value is low for a cryostat that stands 1.5 meters tall with optical access. These losses, along with the pump's flow rate and the pipe head loss, allow the cryostat to reach approximately 1.3 Kelvin at the lowest point.

If the $3.9$~L of the experimental cell is completely filled, it would take 10 hours and 53 minutes to drain it under these experimental conditions. This autonomy is satisfactory, as it is long enough for a full day of work, considering it takes at least 1 hour and 30 minutes to fill all the volumes and prepare the apparatus in the morning. However, this is an overestimate because we clearly cannot study He$_\mathrm{II}$ dynamics when the fluid level is in the middle of the windows. It is best to study He$_\mathrm{II}$ away from its boundaries while maintaining a nearly constant fluid height. Therefore, experiments should be conducted when the superfluid-gas interface is in the upper cylinder of the cell. This part has a volume of $2.84$~L (\diameter 198~mm, 92~mm height), which means its autonomy under experimental conditions is about 8 hours. While the interface remains in this section, its level decreases at a rate of $11.6$~mm per hour. Thus, during a 10-minute movie, the He$_\mathrm{II}$ height drops by approximately 1.93~mm, representing a maximum change of $1.6\%$ in the worst case (liquid level at its lowest). The experimental cell and the reservoir are also equipped with a superconducting gauge whose resistance is monitored by an in-house developed Liquid Helium Level Meter (LHLM) to measure the level of liquid helium.

\subsection{Rotating motion} \label{Rotation}
\subsubsection{Rotating table}
\begin{figure}[hbt!]
    \centering
    \includegraphics[width=0.45\textwidth]{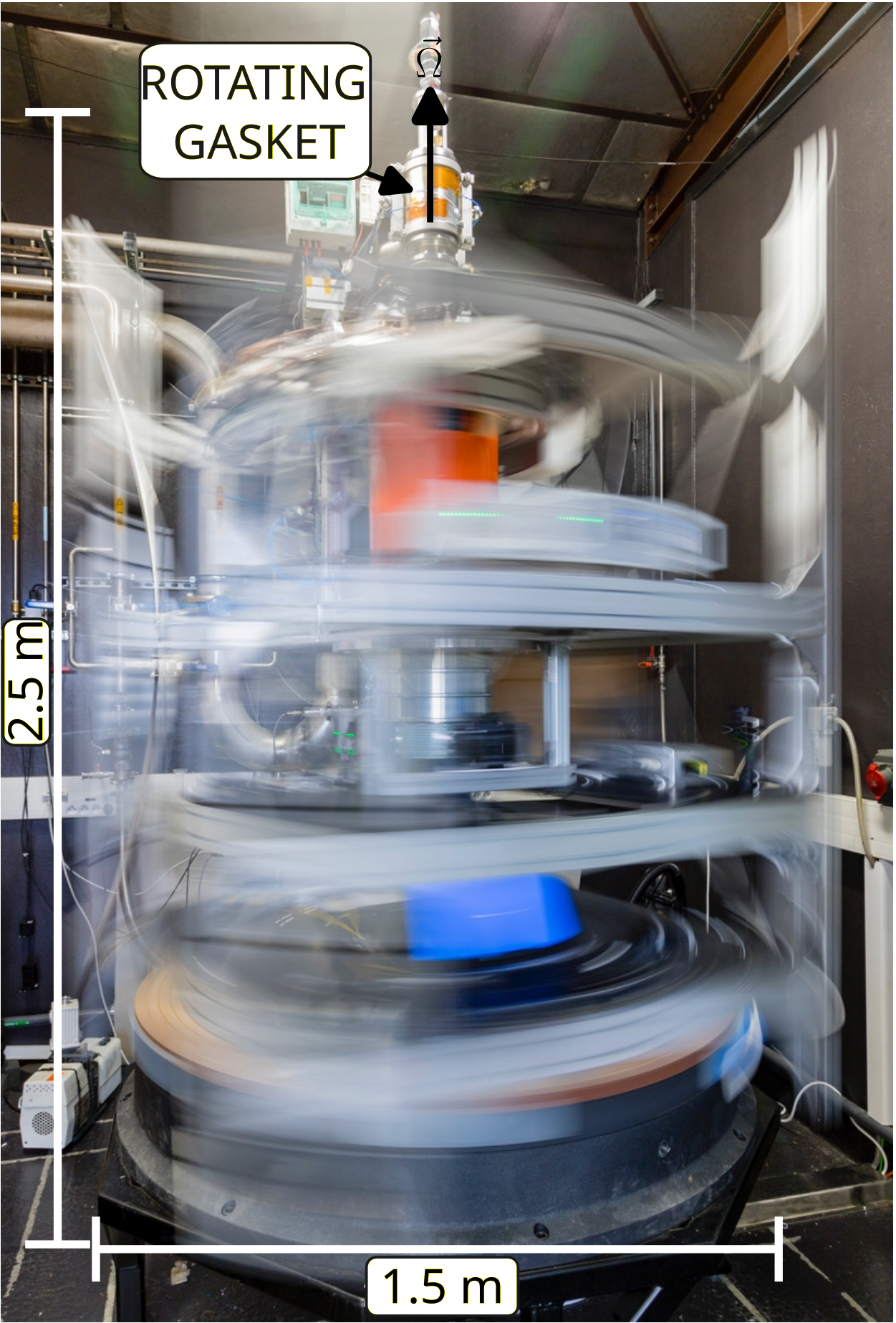}
    \caption{Rotational motion of the entire experimental apparatus is presented. The stators of the rotating table and the rotating gasket are not blurred, as they remain stationary within the laboratory reference frame.}
    \label{tourne_flou}
  \end{figure}

To ensure a stable and smooth rotation free from mechanical vibrations, the CryoLEM is mounted on an air-bearing rotary table. This table, manufactured by Jena Tec, consists of two stacked cast iron rectified disks (\diameter $1.2$~m), each weighing about one ton. A compressed air system in the lower disk allows the upper disk to levitate 50~$\mu$m above the lower one, thereby eliminating dry friction between them. An electric servomotor beneath the table, connected via a transmission belt, enables the stable rotation of the CryoLEM up to $120$~RPM ($2$~Hz) in both directions (see fig.~\ref{tourne_flou}). Most experiments were performed at rotational speeds between $-15$~RPM and $15$~RPM. The relative uncertainty in rotational speed is $2\%$. Speed control is achieved in two ways: using an angular position probe mounted on the electric motor and with accelerometers attached to the rotating disk, which improves the accuracy of $\Omega$. The accelerometers also help detect catastrophic or abnormal events.  

\subsubsection{Alignment of the CryoLEM rotation axis}
\begin{figure}[!htbp]
    \centering
    \includegraphics[width=0.45\textwidth]{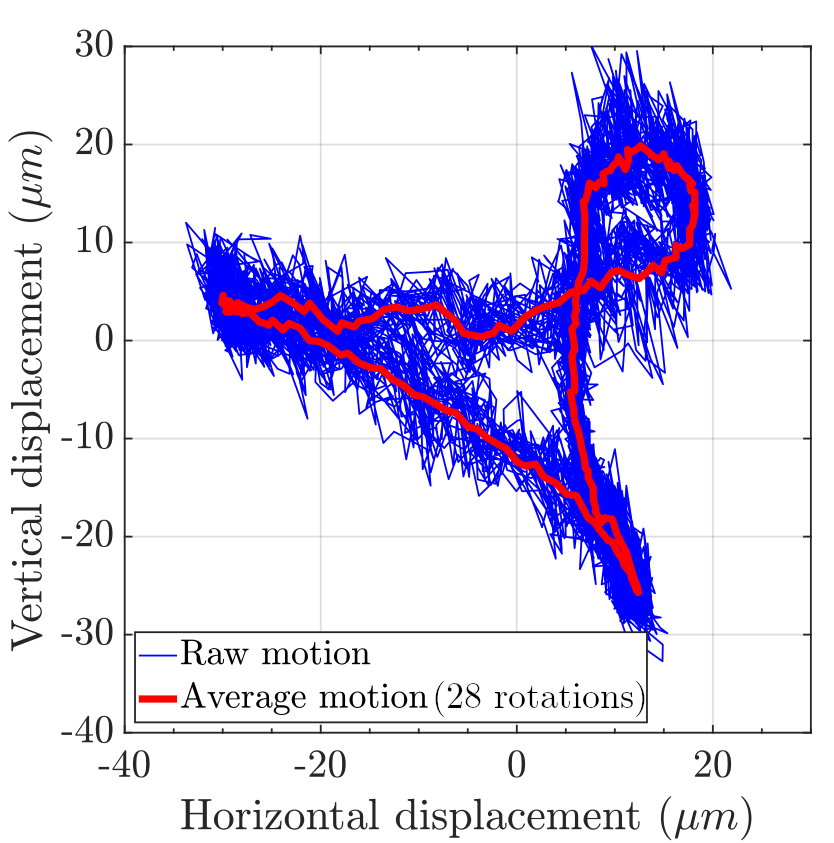}
    \caption{Displacement of the inner cell optical port opposite to the Zyla camera. The window contour is tracked over 28 periods at 15RPM. The experimental cell contains He$_\mathrm{II}$ at $2.08$~K at SVP.}
    \label{accelero}
\end{figure}

The cryostat is placed on the table and aligned using dial gauges attached to the lab walls. Next, the rotation axis of the apparatus and the guiding stator position are optimized by measuring the experimental cell motion relative to the table during rotation. If the guiding stator, located at the top of the cryostat, isn't perfectly aligned with the rotation axis, the experimental cell at the bottom swings like a pendulum when rotating. Due to the asymmetric pressure gradient (the pumping line exits the chimney from the side), a force of $\simeq 2000$~N acts perpendicular to the rotation axis at the top of the cryostat. To counteract this force, adjustable bars are attached to the wall and the top of the cryostat. By changing the lengths of these bars, the guide is adjusted until the cell motion during pumping is minimized. Despite this, we still experience a parasitic periodic mechanical motion of the experimental cell relative to the cameras, with a maximum amplitude 50~$\mu$m, as shown in fig.~\ref{accelero}.

The main cause of this misalignment is that the air bearing of the rotating table allows it to move freely by about a millimeter on each side. This means that displacing mass or pushing it during setup for normal operations could shift the axis, especially if done while the upper disk is levitating. However, the amplitude of the parasitic motion has been consistent across different experiments, indicating that the movement inside the air bearing must be self-aligning to some extent. fig.~\ref{accelero} demonstrates that this motion is both deterministic and periodic based on quantitative evidence.

\subsection{Optical diagnostics - Visualization} \label{Optical diagnostics - Visualisation}
The original goal of this experiment is to do eulerian or lagrangian tracking, therefore 8 optical ports were placed on each shield of the experimental cell. This window configuration makes it possible to study the fluid through any angle (in 2D and 3D).

\subsubsection{Cameras and laser}
Two different cameras were used during our experiments. The first is an Andor Zyla 5.5 sCMOS, capable of recording movies at up to 52 images per second, with a resolution of $2560 \times 2160$, a depth of 16 bits, and a pixel size of $6.5\times 6.5 \ \mathrm{\mu m^2}$. Thanks to a compatible data acquisition card supplied by the manufacturer, images can be transmitted to a computer at bit rates reaching 3.5~Gb/s. The computer, mounted on the rotating structure, is equipped with two SSDs, each with a capacity of 800~GB and 1600~GB. The second camera is a CMOS Phantom v311 by Vision Research, which can record movies at up to 3250 images per second at full resolution ($1280 \times 800$), with a depth of 8 or 12 bits, and a pixel size of $20 \ \mathrm{\mu m}$. Data is written to an internal 8~GB high-speed memory, equivalent to 8000 images at full resolution. Multiple cameras can be mounted simultaneously on the rotating table to record the 3D motion of particles. To capture the motion of micron-sized particles, macro lenses are used. All available lenses are single-focus (50~mm or 100~mm focal length) to minimize distortion. They also feature adjustable apertures, allowing for adjustments to luminosity and depth of field.

\begin{figure}[hbt!]
    \centering
    \includegraphics[width=0.45\textwidth]{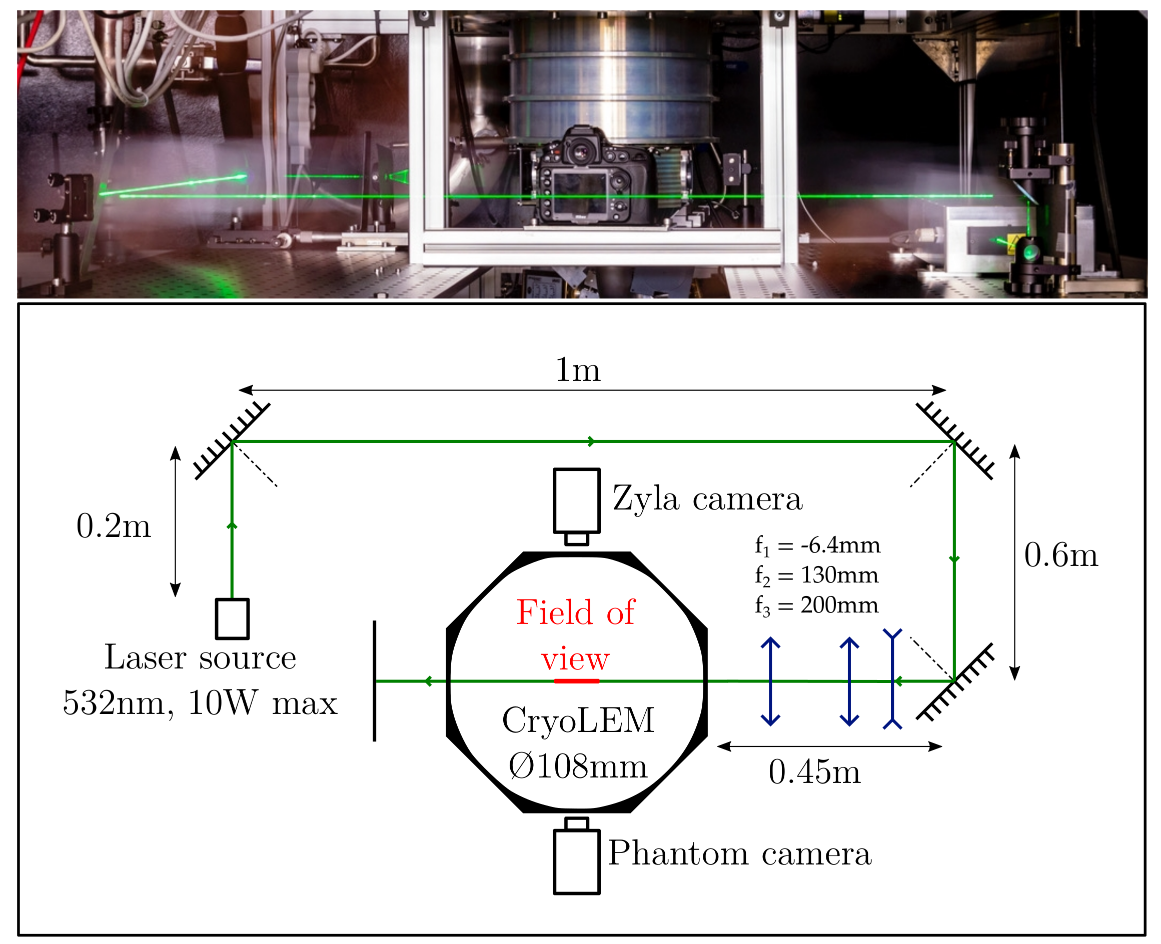}
    \caption{TOP: Artistic photograph showing a side view of the laser beam. The laser source is mounted on the rotating table. BOTTOM: Top-down schematic of the laser beam along the optical setup, in the reference frame of the rotating table.}
    \label{optics}
\end{figure}

An Azurlight ALS-GREEN-10-SF 532~nm continuous fiber laser, with a maximum power of 10~W, is used to produce a laser sheet (see fig.~\ref{optics}). This laser technology can withstand the inertial forces when the table rotates.

\subsubsection{Particles seeding process}
\begin{figure}[hbt!]
    \centering
    \includegraphics[width=0.45\textwidth]{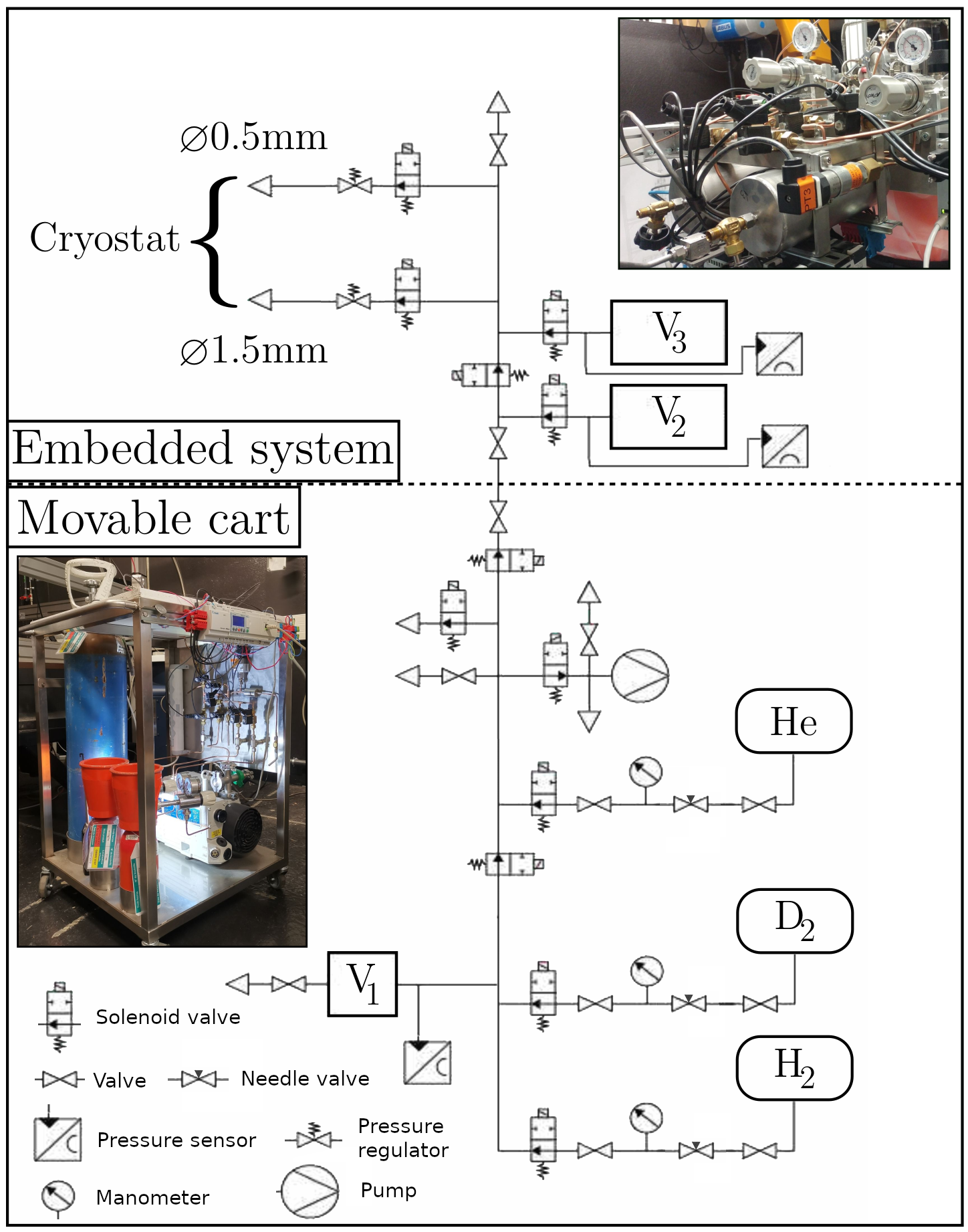}
    \caption{Pneumatic circuit diagram of the particle injection system. The cart includes a pump and bottles of hydrogen, deuterium, and helium. Reservoir V$_1$ is used to prepare the mixture with a precise partial pressure of the seeding component. V$_2$ and V$_3$ contain the gas mixture. All solenoid valves are controlled using a dedicated programmable logic controller.}
    \label{injection}
\end{figure}

To produce particles, two 1~L gas tanks are embedded in the spinning structure. During the preparation of the gaseous mixture, a movable cart connects to the embedded injection system. It includes $^4$He ALPHAGAZ\texttrademark~2, $H_2$, and $D_2$ gas bottles with valves, a pressure control volume, and a pump to manage the feeding of the embedded volumes. Most of the valves on these two pieces of equipment are solenoid valves, enabling remote control. They are operated by two dedicated programmable logic controllers (PLCs) to ensure process reliability and repeatability.
The two embedded tanks, noted as V$_2$ and V$_3$ (see fig.~\ref{injection}), are connected to the liquid helium vessel by two capillaries with inner diameters of \diameter $0.5$~mm and \diameter $1.5$~mm, each approximately $1.5$~m long. These capillaries are made of a cupronickel alloy. By selecting the diameter of the capillary used, it is possible to adjust the injection flow rate, which is a key parameter in particle shaping. Another way to modify the flow rate is by changing the pressure difference between both ends of the injection capillary, as follows:
\begin{equation}
    Q_{inj} = \frac{\pi r_c^4(P_{e}-P_{cryo})}{8\mu L_c}
    \label{injection_formule}
\end{equation}
With $Q_{inj}$ as the expected flow rate, $r_c$ the radius of the capillary, $P_{e}$ the pressure at the capillary entry, $P_{cryo}$ the pressure in the cryostat, $\mu$ is the dynamic viscosity of the mixture flowing through the capillary, and $L_c$ the length of the capillary. Therefore, in addition to the solenoid valve, a pressure regulator was added at the entrance of each injection line. The solenoid valves enable the remote preparation of the mixture and particle injection, while the regulators maintain a constant injection pressure $P_e$. Even after multiple injections that reduce the pressure in reservoirs V$_2$ and V$_3$, the flow rate in the capillaries remains unchanged. Additionally, the injection pressure can be manually adjusted between $0$ and $3$~bar (absolute). This pressure is monitored on the manometer of the pressure regulator, which has a resolution of $50$~mbar. The purpose of the injection system is to produce a cloud of thin, micron-sized particles suitable for PTV and PIV. After numerous tests, we established the following protocol:
\begin{itemize}
    \item Gas mixture composed of $1.4$\% to $2$\% $H_2$ (or $D_2$) completed with He;
    \item Injection just above the He$_\mathrm{II}$/He$_\mathrm{I}$ transition temperature, $T_{\lambda}$;
    \item Using the \diameter$1.5$~mm capillary at a $\Delta P = P_{e}-P_{cryo} \simeq$ $250$~mbar.
\end{itemize}
First, the entire system is pumped down to remove air and any previous gas mixtures. Then, reservoir V$_2$ is filled with deuterium (or hydrogen) to a specific pressure, and topped off with pure helium up to $3.5$~bar. The partial pressure of the seeding gas can be adjusted to alter the mixture ratio, but it is typically set to $52.5$~mbar, which corresponds to $1.5\%$ of seeding gas. Once this mixture is prepared, V$_2$ is closed, and the pipes are pumped again to a vacuum. Reservoir V$_3$ is filled using the same process. Having two embedded tanks V$_2$ and V$_3$ allows us to use two different gas mixtures simultaneously.

\subsection{Protocol} \label{Protocol}

Starting with a cell filled with liquid helium, where the surface level is above the optical ports, the conditions for a new experimental run are considered nominal if $T_i = T_\lambda + \epsilon_T$, $P_i = P_\lambda + \epsilon_P$, with $\epsilon_{T, P}>0$ very small (less than $0.1\mathrm{~K}$ and less than $10\mathrm{~mbar}$ respectively). The CryoLEM is spun long enough to reach solid body rotation, except for a few bubbles, which are unavoidable since the fluid is at saturated vapor pressure. We then proceed as such:

\begin{enumerate}
    \item \label{first} Dihydrogen or dideuterium particles are injected directly into He$_\mathrm{I}$.
    \item Pressure is reduced to just below $T_\lambda$, typically $48$~mbar ($T \simeq 2.157$~K).
    \item The agitation caused by the injection takes a few minutes to subside, and the quick settling of the particles on the vortex lattice is a good indicator of the seeding quality. A video demonstrating the injection is available in Ref.~\onlinecite{videos}. This video also shows the temperature of the helium bath synchronously, precisely illustrating the superfluid transition.
    \item The leak valve connected to the pressure regulator is opened and adjusted to begin pressure reduction. Observing a decrease in pressure causes the particles to rise, likely due to some form of counterflow. Therefore, the pressure decrease should be slow (a few $\mathrm{mbar/min}$) to prevent depleting the particles inside the Region Of Interest (ROI). Once the desired pressure is achieved, the leak valve is closed, maintaining the control volume of the pressure regulator at constant pressure.
    \item With the table spinning, particles injected, laser turned on, and helium bath at the desired temperature, image acquisition can be triggered. At this point, any test can be performed.
    \item Tests and experiments continue until no more particles remain in the ROI as dihydrogen and dideuterium particles respectively float and sink in He$_\mathrm{II}$. Particle depletion occurs in nearly 30 minutes when no heating is applied.
    \item The pressure regulator is isolated, and He$_\mathrm{II}$ evaporation fills the experimental cell with gas, causing the pressure and temperature to rise again within a few minutes. This is maintained just above $T_\lambda$, to revert back to He$_\mathrm{I}$.
    \item The cycle has ended; if time and liquid helium levels allow, we cycle back to \ref{first}. If not, we stop experiments for the day.
\end{enumerate}

Once all experiments are finished, the laser is turned off, rotation stops, and the remaining helium in the cell evaporates. Pressure will rise until matching the recycling line check valve control pressure, ensuring that helium can only escape from the cryostat through the recycling line.

\section{Tools} \label{Tools}
Here we present the tools built around the control of the CryoLEM and the experimental flows in it. 
\subsection{Apparatus control}

\begin{figure*}[!ht]
\centering
    \includegraphics[width=\linewidth]{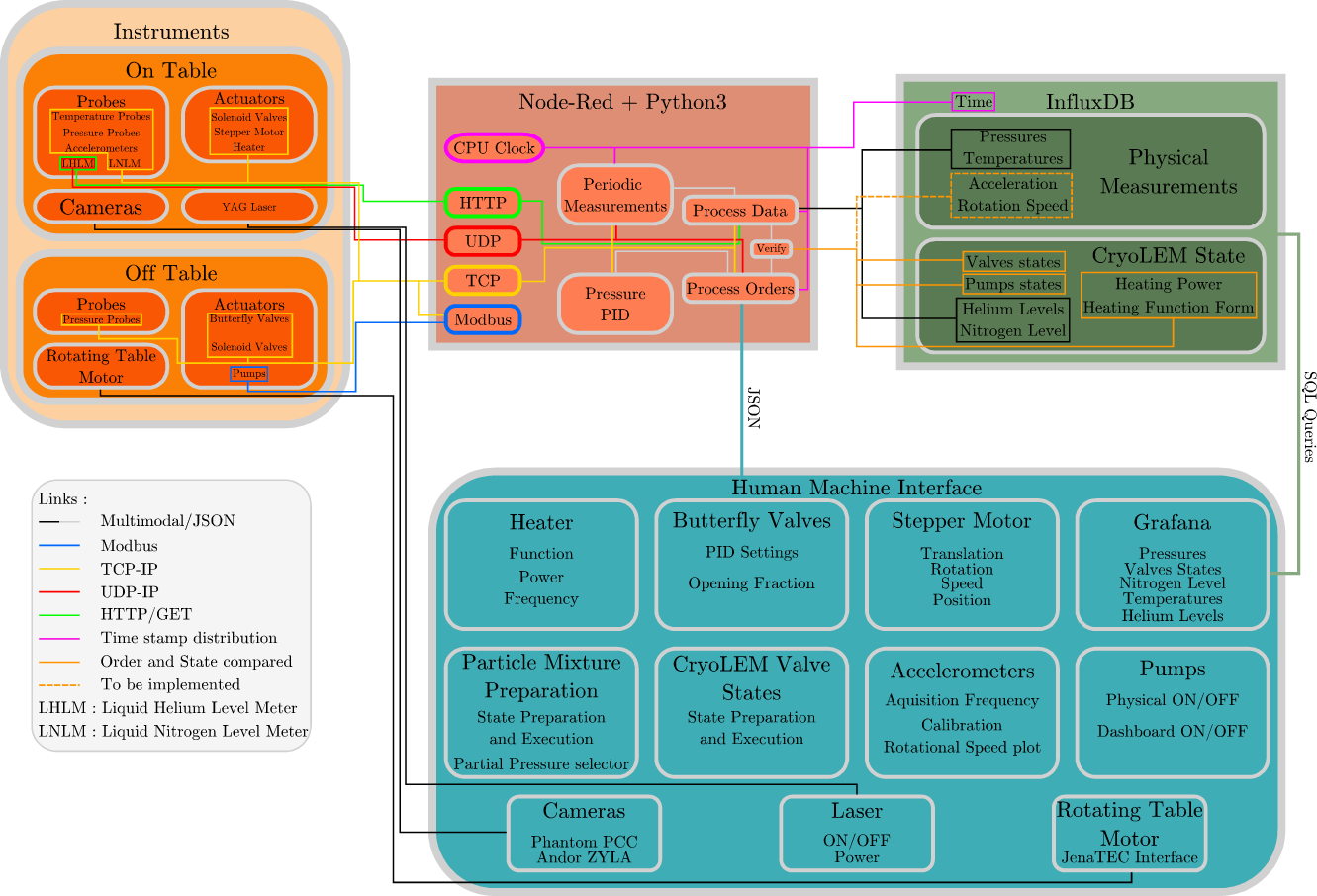}
    \caption{Complete schematic of the CryoLEM acquisition chain, including continuous data logging and remote access control.}
    \label{NR_HMI}
\end{figure*}

Parameter control is essential for reproducible experiments in He$_\mathrm{II}$. Therefore, the sensors and actuators states of the CryoLEM are regularly checked using a combined Node-RED, Python3, and JavaScript algorithm, which logs data into an InfluxDB database. The overall data and control architecture of the CryoLEM is shown in fig.~\ref{NR_HMI}. Because of the rotation, passing information through connecting cables is only practical between sensors and actuators that are embedded in the rotating table. Hence wi-fi is used to remotely actuate and control the experiment from the reference frame of the laboratory. All sensors and actuators are accessible on the Local Area Network.

The CryoLEM relies on a central computer running asynchronous parallel loops that send and receive information from probes and actuators. This exchanged information is logged with individual timestamps, allowing us to synchronize film acquisition with data from the probes. Node-RED makes it easy to create a web page accessible on the local network, featuring a control interface with buttons, text boxes, sliders, drop-down menus, simple graphs, and clickable SVGs. For analyzing previously acquired data, in addition to films or photos, the use of Grafana to visualize the logged data in InfluxDB has greatly improved our understanding of the CryoLEM. Ultimately, this enabled us to develop multiple dashboards, each tailored to specific steps in our protocols, thereby reducing the risk of human errors.

\subsection{Stepper motor}

A stepper motor is mounted at the top of the cryostat on the spinning axis, with its shaft extending through the entire cryostat chimney (see fig.~\ref{SchemaCryostat}) down to the experimental helium bath. The shaft measures 1.45~m in length and passes through several guides to prevent flexing and ensure rotational stability. It is hollow, which reduces the torque required to drive it and minimizes heat transfer by conduction through its structure.
The motor is a Haydon Kerk - Ametek LR35MH4AE-12-840. It features two independent axes: one for rotation and one for translation, with a maximum stroke of $84$~mm. The rotating axis has 200 steps per revolution, and the translating one has 25 steps per mm. Both feature adjustable microsteps, which means that motor steps can be divided by a value between 2 and 28, for improved precision (but lower velocities). In the current configuration, speeds of the order 200~RPM in rotation, and 10~mm.s$^{-1}$ in translation can be achieved.
This enhances the modularity of the CryoLEM, allowing different fluid forcing methods without altering the entire mechanical power chain. Initially, a flat propeller was mounted to stir the fluid and homogenize the particle field. In this setup, the propeller is translated close to the field of view, rotated to stir the particles, and then moved as far back as possible to cancel out its influence on the studied flow. This translation could also be used in future experiments on grid turbulence (decaying turbulence if pulled once \cite{Stalp,Smith,cortet}, or steady state if oscillated \cite{Sy,Sy_article,Mantia}, by replacing the propeller at the shaft tip. Another advantage would be calibrating 3D visualization techniques: a test pattern fixed at the shaft tip can be moved within the field of view to calibrate the cameras.
The driver for these motors is a Trinamic TMCM-3110. This board is connected through the RS485 port to a Brainbox\textregistered{} ES-313, which serves as a serial-to-Ethernet adapter. The Brainbox\textregistered{} links to the lab network, enabling remote communication with the main computer. The driver uses software called TMCL-IDE 3.0, which runs on a computer to communicate with the motor and program its movements. It allows different settings to be converted into variable-length hexadecimal messages sent to the driver, which controls the motor axes. This software also includes a programmable section where users can write scripts with operations like loops. For example, the motor can be programmed to move up and down at a constant velocity (trapezoidal velocity signal) to create an oscillating motion, useful for studying grid turbulence. The software can also display graphs of the motor position and velocity.
Until now, the motor has been used only for simple applications, such as moving an object of known size into the field of view to calibrate the cameras or for continuous rotation of a propeller to study steady rotation or counter-rotating turbulence. To centralize as many devices as possible into a single interface, a Python3 code was written to control the motor. 
This code allows the user to set all the motor parameters. The Python3 code translates the user’s commands into the corresponding hexadecimal message, which is then sent to the motor driver. A variety of use cases are presented in E. Durozoy's thesis \cite{Emeric}.

\section{Conclusion and perspectives}
The validation of tracking quantum vortices in He$_\mathrm{II}$ with solid particles of dihydrogen or dideuterium was conducted using a rotating canonical flow \cite{Peretti}. This experiment shows that the CryoLEM is a reliable platform, capable of supporting additional experiments, especially since rotation combined with temperature control offers a reproducible initial condition for future studies. One such experiment is the rotating counterflow, examined by Swanson et al. \cite{Swanson} and Dwivedi et al. \cite{Varga} and qualitatively studied by Peretti et al.\cite{Peretti}. This phenomenon, unique to He$_\mathrm{II}$, exists within the three-dimensional parameter space of $\dot{Q}$ the heating power, $T$ the temperature, and $\Omega$ the rotating speed, all of which are controlled in the CryoLEM. 

\begin{figure}[!htbp]
    \centering
    \includegraphics[width=0.45\textwidth]{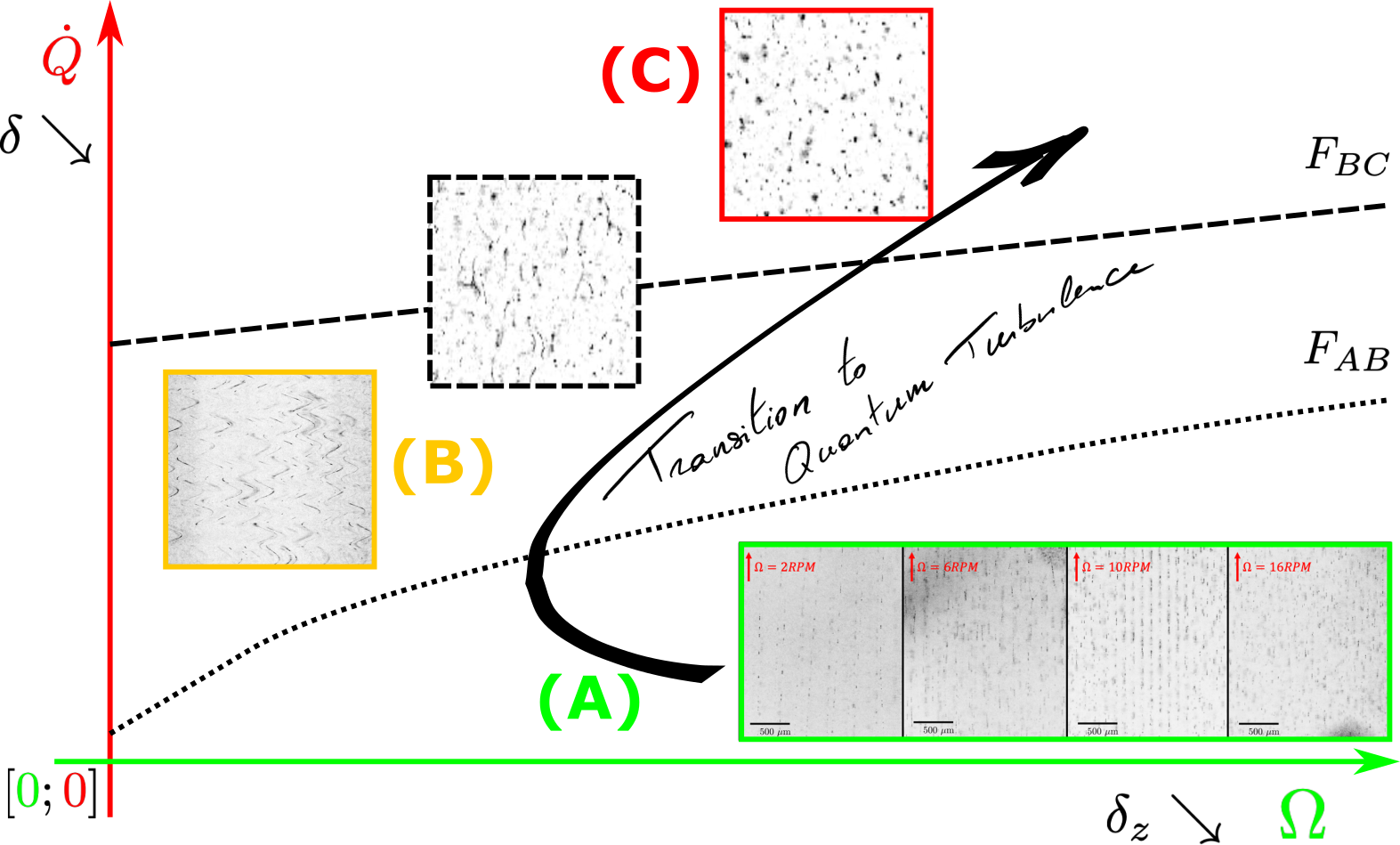}
    \caption{Rotational velocity $\Omega$ and heating power $\dot{Q} \ (\equiv v_{ns})$ phase space. Three regimes are visually observed during experiments. The first regime (A) is the stable canonical pure rotation vortex lattice. Regimes (B) and (C) are counterflow under rotation, with the key difference that regime (B) shows wave propagation along the vortex lines, while regime (C) is fully turbulent flow. }
    \label{Phasespace}
\end{figure}

\begin{figure}[!htp]
    \centering
    \includegraphics[width=0.6\linewidth]{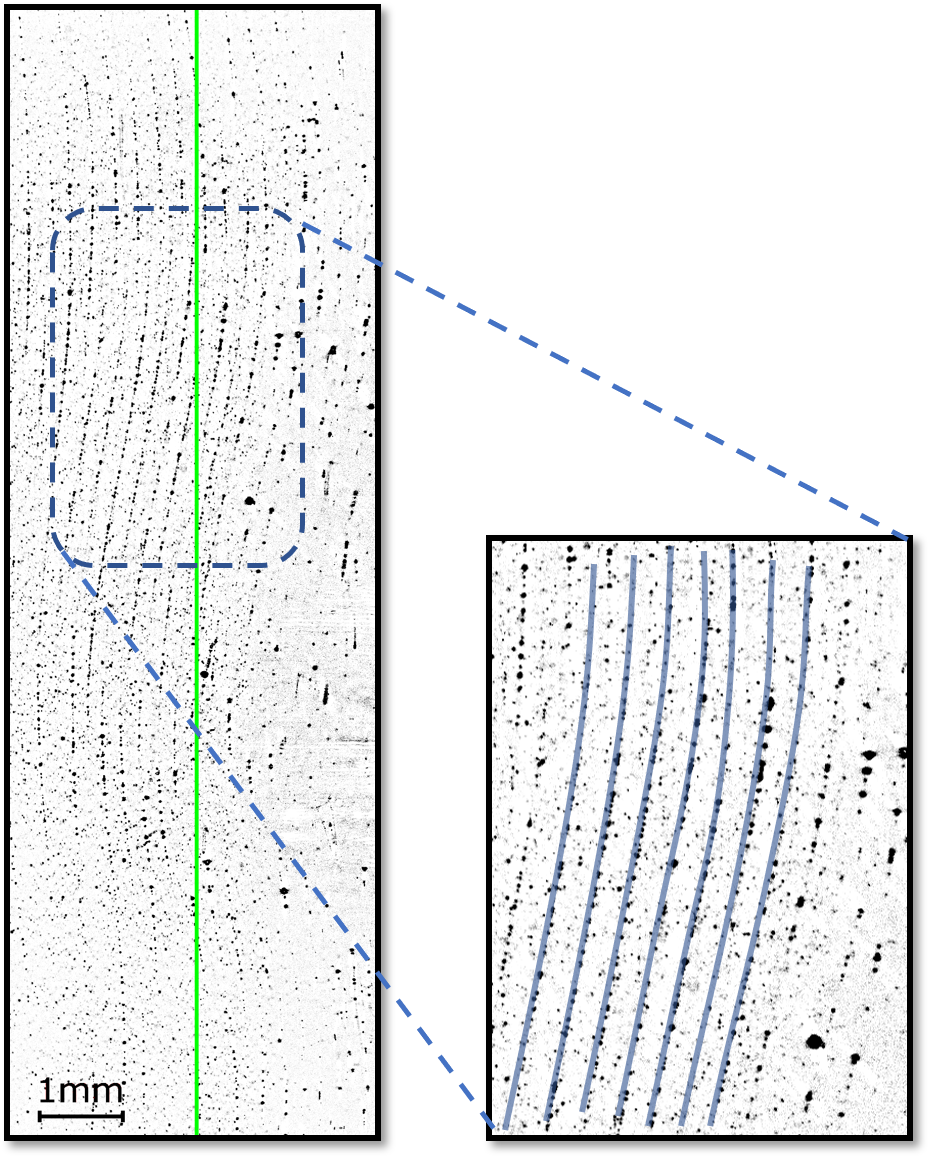}
    \caption{Wavelike perturbation on a vortex lattice. In the inset, the vortices are highlighted for clarity. The video is available for download in this Ref.~\onlinecite{videos}}
    \label{fig:waves}
\end{figure}

Rotating counterflow exploration reveals different regimes (see fig.~\ref{Phasespace}), such as waves (see fig.~\ref{fig:waves} and the second downloadable video Ref.~\onlinecite{videos}, corresponding to the (B) region in fig.~\ref{Phasespace}. This regime, characterized by continuous lattice perturbation, occurs when the heat flux remains moderate. The fact that the waves seem to propagate as a collective mode on the lattice is a novel feature, as far as we know, since other observations of Kelvin waves in He$_\mathrm{II}$ \cite{fonda2014kelvin,minowa} result from local perturbations on individual vortices. Additionally, global excitation of Kelvin waves can also induce turbulence, as experimentally observed in the $^3$He-B superfluid phase \cite{Makinen}. Interactions among inertial forces, vortices, and the normal fluid are believed to play a role in this process. However, their exact nature remains unclear and will be investigated in future work.

Furthermore, the CryoLEM is not limited by its technical capabilities. Its large experimental volume allows for the operation of Eulerian sensors \cite{diribarne2021investigation,salort2021experimental} outside the region of interest in parallel with direct visualization, enabling comparisons between new visual results and local measurements. Additionally, various mechanical setups can be mounted on the end of the rotor axis, supporting a wide range of experiments similar to classical flows, with the added benefit that studies can be carried out in both He$_\mathrm{I}$ and He$_\mathrm{II}$. Ultimately, the CryoLEM, as well as "similar experiments on the generation of QT by perturbing a rotating vortex lattice with well-known structure, representing well-defined starting conditions for the generation of waves or 3D QT [...] should shed new light on the generation of QT in helium superfluids" (as stated by Skrbek \textit{et al.} \cite{sreeni_review}) and its underlying phenomena, such as Kelvin waves or quantum vortex reconnection.

\nocite{*}

\section{Acknowledgments}
This work received the support of grants ANR-11-PDOC-0001 (3D-QuantumV), ANR-10-LABX-0051 (LANEF), ANR-17-CE30-0003 (DisET) and ANR-23-CE30-0024 (QuantumVIW).

\section{References}
\bibliography{main}

\end{document}